\documentclass[a4paper,10pt]{article}
\usepackage[utf8x]{inputenc}
\usepackage{amssymb,amsfonts,amstext,amsmath,graphicx,epic,amsthm,color,times}

\title{{Recherche et \'etude de la stabilit\'e du cycle limite pour l'oscillateur de Rayleigh}
}
\author{C. H. Miwadinou\footnote{clement.miwadinou@imsp-uac.org}, A. V. Monwanou\footnote{movins2008@yahoo.fr} and J. B. Chabi
 Orou\footnote{Author to whom correspondence should be addressed: jchabi@yahoo.fr}}

\begin{document}

\maketitle Institut de Math\'ematiques et de Sciences Physiques, BP: 613 Porto Novo, B\'enin

\begin{abstract}
Dans ce papier, nous avons étudié l'oscillateur autonome de Rayleigh  en cherchant son cycle limite et en étudiant la stabilité de ce cycle. 
A travers cette étude, nous avons étudié les points fixes de l'équation de l'oscillateur de Rayleigh et nous nous sommes rendus compte qu'en réalité 
cet oscillateur n'a qu'un seul point fixe. Ce qui paraît nouveau de notre point de vue,est que nous avons trouvé le cycle limite de cet oscillateur en utilisant l'une  des formes  du théorème de
 Poincaré-Bendixson puis nous avons étudié la stabilité de ce cycle limite par la méthode des échelles multiples. Enfin, le plus important dans ce papier 
est que nous avons prouvé analytiquement puis confirmé par la simulation numérique en utilisant Mathématica que l'oscillateur de Rayleigh présente une bifurcation de PAH   
\end{abstract}

{\bf Mots clés}: Oscillateur de Rayleigh, point fixe, cycle limite, stabilité, bifurcation.

\section{Introduction}
L'étude des oscillateurs non-lin\'{e}aires est une th\'{e}matique actuellement importante, tant d'un point de vue acad\'{e}mique 
qu'industrielle et qui touche de nombreux domaines, tels que l'hydrodynamique, l'a\'{e}ronautique, le g\'{e}nie civil, les transports, l'acoustique
 musicale, le g\'{e}nie nucl\'{e}aire et autres. Les premiers travaux datent du $XIX^e$ si\`{e}cle
  mais actuellement ils conna\^{i}ssent un regain d'int\'{e}r\^{e}t du fait du besoin d'optimiser, d'all\'{e}ger les structures
 couramment utilis\'{e}es et soumises \`a des niveaux d'excitation importante, ou de contr\^{o}ler les instabilit\'{e}s des oscillations. Dans ce sens, 
l'étude des oscillateurs auto-excités paraît très compliquée en ce qui conserne la recherche de leurs solutions analytiques. Pour y arriver, plusieurs 
techniques ont été proposées pour la détermination des solutions approchées de ces système.
Lord Rayleigh, au cours de ses travaux dans les ann\'ees 1887 sur la description du fonctionnement des instruments de musique, 
a d\'ecouvert une cat\'egorie d'oscillateurs qui portent son nom et qui permettent aussi aujourd'hui de modéliser plusieurs systèmes surtout en ingénierie.
 C'est ainsi que dans l'une de ses oeuvres [1], il pr\'esente plusieurs 
mod\`eles pour d\'ecrire la vibration de diff\'erents instruments de musique. Malgr\'e la non- lin\'earit\'e des oscillateurs consid\'er\'es, 
il trouve pour certains d'entre eux des cycles limites comme solution du mouvement. Ces cycles limites sont atteints par les oscillateurs, 
peu importe la condition initiale. La mod\'elisation de ces oscillateurs est donn\'ee par l'\'equation de Rayleigh suivante:
\begin{equation}
\ddot{X}+X+\epsilon f(X,\dot{X})\dot{X}=0 
\end{equation}
Par utilisation d'un temps r\'eduit, la fonction d'amortissement f r\'eduite qu'il a propos\'ee est:
\begin{equation*}
 f(X,\dot{X})= -1+\dot{X}^2 
\end{equation*}
Ainsi, l'\'equation réduite de l'oscillateur de Rayleigh est la suivante
\begin{equation}
\ddot{X}+X= \epsilon (1-\dot{X}^2)\dot{X}
\end{equation}  
o\`u $\epsilon$ est un param\`etre de bifurcation.

En 2000, Johan a étudié cet oscillateur en étudiant ses points fixes. Il a utilisé le théorème de Poincaré-Bendixson pour prouver l'existence du 
cycle limite 
de l'oscillateur autonome de Rayleigh puis il a étudié la stabilité de ce cycle limite en utilisant la technique de Poincaré-Bendixson. Dans notre 
travail, nous allons 
rechercher le cycle limite de l'oscillateur autonome de Rayleigh par une autre variante du théorème de Poincaré-Bendixson portant sur un système 
hamiltonien plan perturbé.
Nous allons dans un deuxième temps étudier la stabilité du cycle limite trouvé en utilisant la méthode générale dite d'échelles multiples qui selon 
les littératures traite
 les oscillateurs dans toutes les conditions. Troisièmement, nous allons prouver que l'oscillateur autonome de Rayleigh présente une bifurcation de PAH. 
Enfin nous allons simuler l'équation de cet oscillateur pour vérifier nos résultats.

\section{Points fixes et existence de cycle limite de l'oscillateur de Rayleigh}

\subsection{Etude des points fixes}
Pour effectuer une \'etude g\'en\'erale de cet oscillateur, mettons  \'equation (2)  sous forme de syst\`eme d'\'equations diff\'erentielles d'ordre 1.\\
En effet, posons $\dot{X}=Y$. On a $ \ddot{X}=\dot{Y}=-X+\epsilon(1-Y^2)Y $\\
On obtient:
 \begin{equation}
 \left\{
              \begin{array}{rl} 
            &  \dot{X}=Y \\
             & \dot{Y}=-X+\epsilon(1-Y^2)Y     
                 
                      \end{array}   
                         \right.     
   \end{equation}                

 Lorsque nous  cherchons les points fixes de ce syst\`eme dynamique, nous constatons que, l'origine c'est-\`a dire $X=0$ et $Y=0$ est le seul point fixe de ce syst\`eme dynamique.\\
Nous cherchons la stabilit\'e de point fixe en étudiant la nature des valeurs propres de la matrice jacobienne de ce syst\`eme dynamique. 

Cette matrice jacobienne est d\'efinie à l'origine par: 
                
 $$
{\mathcal M =}
\left(
\begin{array}{cc}
   0 & 1  \\
  -1 &\epsilon 
\end{array}
\right)
  $$                   

Les valeurs propres $\lambda $ de la jacobienne sont donn\'ees par
\begin{equation}
 \lambda^2-\epsilon\lambda+1=0
\end{equation}
 \underline{Discussion}

$\ast$ Si $|\epsilon|>2,  \triangle>0$ et la jacobienne admet deux valeurs propres $\lambda_1=\frac{\epsilon-\sqrt{\epsilon^2-4}}{2}$ et $\lambda_2=\frac{\epsilon+\sqrt{\epsilon^2-4}}{2}$\\
 Nous prouvons que $\lambda_1$ et $\lambda_2$ sont de m\^{e}me signe. Ainsi,

$\bullet$ Si $\epsilon<-2$, alors $\lambda_1$ et $\lambda_2$ sont n\'egatives. Donc l'origine est un noeud attractif et est stable.

$\bullet$ Si $\epsilon>2$, alors $\lambda_1$ et $\lambda_2$ sont positives. Par suite l'origine est un noeud r\'epulsif.

$\ast$ Si $\epsilon=2$ ou $\epsilon=-2, \triangle=0 $ et donc $\lambda_1=\lambda_2=\frac{\epsilon}{2}$.\\
Il vient:

 $\bullet \lambda_1$ et $\lambda_2$ n\'egatives pour $\epsilon=-2$ et donc l'origine est un noeud attractif.

 $\bullet$ $\lambda_1$ et $\lambda_2$ positives pour $\epsilon=2$ et donc l'origine est un point noeud r\'epulsif.

     $*$ Etudions maintenant le cas $|\epsilon|<2 $\\
 $|\epsilon|<2   \Longrightarrow \triangle<0 $ et $\lambda_\pm=\frac{\epsilon\pm i\sqrt{4-\epsilon^2}}{2} $ \\
Les valeurs propres sont complexes donc l'origine est un foyer ou un centre.

$\bullet$ De plus, pour $-2<\epsilon<0, Re(\lambda_+)=Re(\lambda_-)<0 $ dans ce cas, l'origine est un foyer attractif.

$\bullet $ Pour $0<\epsilon<2, Re(\lambda_+)=Re(\lambda_-)>0 $ donc l'origine dans ce cas est un foyer r\'epulsif.

$\ast$  Enfin si $\epsilon=0, \lambda_{\pm}=\pm2i$, alors l'origine est un centre.

\subsection{Existence d'orbite p\'eriodique}

Dans cette partie, nous allons appliquer le th\'eor\`eme de Poincar\'e-Bendixson énoncé dans le cas des perturbations autonomes 
d'un système hamiltonien plan (J. BRICMONT [2]).

L'équation (2) de l'oscillateur de Rayleigh est un système plan perturbé.Nous considérons donc le système

%
%
%

\begin{equation}
 \left\{
              \begin{array}{rl} 
            &  \dot{X}=Y \\
             & \dot{Y}=-X+\epsilon(1-Y^2)Y     
                 
                      \end{array}   
                         \right.             
\end{equation}
Le syst\`eme non perturb\'e (oscillateur harmonique) définie par $\epsilon=0$ est hamiltonien dont l'hamiltonien est $ H=\frac{1}{2}(X^2+Y^2)$.
 Hormis l'\'equilibre (0,0), les solutions de ce système hamiltonien sont toutes p\'eriodiques de p\'eriode minimale $T_r= 2\pi$ et donn\'ees par 
\begin{equation}
  \Gamma_r:\overrightarrow {X}(t,r)=(r\cos t, -r\sin t)^\top, r>0 
\end{equation} 
Nous calculons  l'int\'egrale de Melnikov le long de $\Gamma_r$
\begin{equation}
  M(r)=\int_0^{T_r} f(\overrightarrow {X}(t,r))\wedge g(\overrightarrow {X}(t,r))dt
\end{equation}
 avec $ f(\overrightarrow {X}(t,r))=(Y,-X)^\top$ et $g(\overrightarrow {X}(t,r))=(0,Y-Y^3)^\top $  
où f et g sont de classe $C^\infty$
\begin{eqnarray*}
M(r)&=&
\int_0^{2\pi}(r^2\sin^2t-r^4\sin^4t)dt 
\end{eqnarray*}
Nous trouvons après calcul que
\begin{equation}
M(r)=\frac{1}{4}\pi r^2(4-3r^2)
\end{equation}

Remarquons que $\sqrt{\frac{4}{3}}$ est la seule valeur de r vérifiant $M(\sqrt{\frac{4}{3}})=0$ et $M^\prime(\sqrt{\frac{4}{3}})=-2\pi\sqrt{\frac{4}{3}}\neq0$.
Il s'en suit donc que toutes les hypothèses du  th\'eor\`eme de Poincar\'e-Bendixson énoncé dans le cas des perturbations autonomes 
d'un système hamiltonien plan sont vérifiées par l'oscillateur de Rayleigh. Nous en d\'eduisons  que l'oscillateur de Rayleigh poss\`ede pour $\epsilon$ suffisamment petit, un cycle limite qui tend vers le cercle dont le centre est l'origne et de rayon$\sqrt{\frac{4}{3}}$ quand $\epsilon$ tend vers 0.

\section{Etude de la stabilit\'e du cycle limite}

Dans cette section, nous allons utiliser la méthode des échelles multiples pour étudier la stabilité du cycle limite de l'oscillateur de Rayleigh.
 Signalons qu'elle n'est pas la seule méthode. Mais ceci suggère de faire reposer la méthode la plus générale cherchée sur une hiérarchie
d’échelles de temps aptes à rectifier le défaut de synchronisation entre l’oscillateur linéaire de référence et
l’oscillateur non-linéaire considéré. Ainsi le but de cette méthode est bien de rendre l’approximation de la solution
uniforme en temps, un peu comme on corrige un défaut de calendrier par l’introduction des années bissextiles
suivant un algorithme d’autant plus compliquée que l’on considère de longues périodes Paul Manneville[3].

Définissons les formules de base de la m\'ethode des \'echelles multiples.
\begin{equation}
 \frac{d}{dt}=\partial_{t_0}+\epsilon\partial_{t_1}+\epsilon^2\partial_{t_2}+.....      
\end{equation}
\begin{equation}
 \frac{d^2}{dt^2}=\partial_{t_0^2}+2\epsilon\partial_{t_0}\partial_{t_1}+\epsilon^2(\partial_{t_1^2}+2\partial_{t_0}\partial_{t_2})+....
\end{equation}
   \begin{equation}
     X=X_0(t_0,t_1,t_2,...)+\epsilon X_1(t_0,t_1,t_2,...)+.... 
   \end{equation}
$t_0=t, t_1=\epsilon t,t_2=\epsilon^2t $,... et $\frac{\partial}{\partial t_i}=\partial_{t_i} $

     Ainsi, lorsque nous remplaçons (9), (10) et (11) dans (2), nous trouvons:
\begin{eqnarray}
[\partial_{t_0^2}+2\epsilon\partial_{t_0}\partial_{t_1}+
\epsilon^2(\partial_{t_1^2}+2\partial_{t_0}\partial_{t_2})+....](X_0+\epsilon X_1+\epsilon^2X_2+....)\cr
+(X_0+\epsilon X_1+\epsilon^2X_2+....)=\epsilon[1-(\partial_{t_0}X_0)^2-
2\epsilon(\partial_{t_0}X_0)(\partial_{t_0}X_1+\partial_{t_1}X_0)\cr+...][\partial_{t_0}+
\epsilon\partial_{t_1}+\epsilon^2\partial_{t_2}+.....](X_0+\epsilon X_1+\epsilon^2X_2+....) 
\end{eqnarray}
Nous obtenons:

A l'ordere $\epsilon^0$
\begin{equation}
\partial_{t_0^2}X_0+X_0=0 
\end{equation}
A l'ordre $\epsilon^1$

\begin{equation}
 \partial_{t_0^2}X_1+X_1=[1-(\partial_{t_0}X_0)^2]\partial_{t_0}X_0 -2\partial_{t_0}\partial_{t_1}X_0 
\end{equation}
Les solutions de l'équation (13)
\begin{equation}
 X_0=A_0(t_1,t_2,...)\cos[t_0+\phi_0(t_1,t_2,...)] 
\end{equation}

En utilisant (15), (14) devient
\begin{eqnarray}
 \partial_{t_0^2}X_1+X_1=[-A_0+\frac{3}{4}A_0^3+2(\partial_{t_1}A_0)]\sin(t_0+\phi_0)+\cr
2A_0(\partial_{t_1}\phi_0)\cos(t_0+\phi_0)-\frac{1}{4}A_0^3\sin3(t_0+\phi_0) 
\end{eqnarray}
Annulons les termes r\'esonants, donc on a:
\begin{equation}
 -A_0+\frac{3}{4}A_0^3+2\partial_{t_1}A_0=0
\end{equation}
\begin{equation}
 2A_0\partial_{t_1}\phi_0 =0\label{eq1}
\end{equation}

L'intégration de l'équation (17) donne
\begin{equation}
 \phi_0=\phi_0(t_2,t_3,...)
\end{equation}
  Lorsque nous passons \`a t \`a l'aide de $t_1=\epsilon t$, l'équation (18) devient: 
\begin{equation}
 \frac{dA_0}{dt}=\frac{1}{2}\epsilon A_0(1-\frac{3}{4}A_0^2)
\end{equation}

Il en r\'esulte qu'il n'y a pas de correction \`a la phase et que la phase est fonction du temps.Nous \'etudions la stabilit\'e du cycle limite en utilisant l'équation (20). En effet, l'amplitude du cycle limite  est 
\begin{equation}
 A_{0S}=\sqrt{\frac{4}{3}}
\end{equation}
La stabilit\'e de ce cycle limite se voit \`a travers le signe de $\frac{dA_0}{dt}$.\\
Ainsi pour $ (1-\frac{3}{4}A_0^2)<0$ c'est-\`a dire $A_0^2>A_{0s}^2, \frac{ dA_0}{dt}<0 $ donc toute orbite d'amplitude sup\'erieure \`a celle du cycle limite chute toujours vers ce dernier. Pour   $ (1-\frac{3}{4}A_0^2)>0$ c'est-\`a dire $A_0^2<A_{0s}^2, \frac{dA_0}{dt}>0 $ donc toute orbite caract\'eris\'ee par une amplitude inf\'erieure \`a celle du cycle limite stable va mourir sur ce dernier.\\
Par cons\'equent, le cycle limite centr\'e sur l'origine d'amplitude $A_{0S}=\sqrt{\frac{4}{3}}$ est stable. Il faut remarquer que ce r\'esultat est le m\^{e}me quelle que soit la m\'ethode de calcul perturbatif utilis\'ee.

\section{Recherche de la bifurcation de Poincar\'e-Andronov-Hopf pour l'oscillateur de Rayleigh}

La bifurcation de Poincar\'e-Andronov-Hopf(PAH) courrament appel\'ee bifurcation de Hopf (parce qu'elle est la forme simple de la bifurcation de Hopf) est l'apparition ou la disparition d'une orbite p\'eriodique par un changement local de propri\'et\'e de stabilit\'e d'un point fixe d'un syst\`eme dynamique. Cette naissance locale ou mort d'une solution p\'eriodique d'oscillation auto-excit\'ee d'un point d'\'equilibre se produit lorsqu'un param\`etre de contr\^ole traverse une valeur critique.
En g\'en\'eral, dans une \'equation diff\'erentielle, une bifurcation de Hopf se produit quand une paire de valeurs propres complexes conjug\'ees du syst\`eme lin\'earis\'e \`a un point fixe devient purement imaginaire. Cela implique qu'une bifurcation de Hopf ne peut se produit dans des syst\`emes de dimensions deux ou plus( en dimension deux, il s'agit d'une bifurcation de PAH).
La recherche de la bifurcation de Hopf en g\'en\'eral et de PAH en particulier n'est pas \'equivalente \`a la recherche de cycles limites stables. D'abord, quelques bifurcations de Hopf( par exemple, sous-critique) n'impliquent pas l'existence de cycles limites stables; deuxi\`emement, il peut exister des cycles limites non li\'es \`a des bifurcations de Hopf.  

Dans cette section, nous allons appliquer le théorème de Poincar\'e-Andronov-Hopf (PAH) \cite{4}. Pour cette raison, nous allons chercher à vérifier si l'oscillateur de Rayleigh remplir les hypothèses de ce théorème. 
 Consid\'erons l'\'equation suivante de l'oscillateur de Rayleigh:

 $ \ddot{X}+X=\epsilon(1-\dot{X}^2)\dot{X} $  o\`u $ \epsilon $ est un param\`etre de contr\^{o}le\\
Nous avons vu ci-haut que cette \'equation s'\'ecrit par d\'ecomposition comme le système (3).
Ainsi, ce syst\`eme est sous forme 

 \begin{equation}
 \left\{
              \begin{array}{rl} 
            &  \dot{X}=f(\epsilon,X,Y) \\
             & \dot{Y}=g(\epsilon,X,Y)    
                 
                      \end{array}   
                         \right.     
   \end{equation}               

avec $ f(\epsilon,X,Y)=Y$ et $ g(\epsilon,X,Y)=-X+\epsilon(1-Y^2)Y $

1) Les fonctions f et g sont des polyn\^{o}mes donc elles sont diff\'erentiables de classe $C^\infty $\\
2) Pour $X=0$ et $Y=0$, $f(\epsilon,0,0)=0 $ et $ g(\epsilon,0,0)=0 $\\
3) D'apr\`es l'\'etude des points fixes de l'oscillateur de Rayleigh faite dans la sous-section 2.1, nous avons pour $| \epsilon |<2$ deux racines complexes conjugu\'ees de la jacobienne du syst\`eme dynamique (3) d\'efini  \`a l'origine qui est le seul point fixe de ce syst\`eme. Ces valeurs propres sont:\\
$\lambda\pm=\frac{\epsilon\pm i\sqrt{4-\epsilon^2}}{2}$ donc $\lambda\pm =\alpha(\epsilon)\pm i\beta(\epsilon) $ avec $\alpha(\epsilon)=\frac{1}{2}\epsilon$ et $\beta(\epsilon)=\frac{\sqrt{4-\epsilon^2}}{2} $.\\
Ainsi, on a:\\
$ \epsilon<0 \Longrightarrow \alpha(\epsilon)<0, \epsilon>0 \Longrightarrow \alpha(\epsilon)>0,  \alpha\prime(0)=\frac{d\alpha}{d\epsilon}=\frac{1}{2}\neq 0 $ et $\beta(0)=1\neq 0 $\\
4) Toujours dans la sous-section 2.1, nous avons montr\'e que l'origine est un centre et d'apr\`es l'\'etude de la stabilit\'e, l'origine est un  attracteur.

Ces quatre points montrent que les quatre hypothèses du théorème de PAH sont satisfaites par l'équation de l'oscillateur de Rayleigh.

   En vertu du th\'eor\`eme de PAH, l'oscillateur de Rayleigh est tel que:\\
- Pour $\epsilon<0 $, l'origine est un \'etat stationnaire stable.\\
- Pour $\epsilon>0  $, l'origine est un \'etat stationnaire instable entour\'ee par un cycle limite stable dont le diam\`etre est de l'ordre de$\sqrt{\epsilon}$.\\
Il vient que $\epsilon=0 $ est une valeur de bifurcation de l'oscillateur de Rayleigh.\\
Autrement dit, lorsque $\epsilon $ passe par 0, l'origine perd sa stabilit\'e. Cette perte de stabilit\'e s'accompagne de la naissance d'un cycle limite stable dont le rayon cro\^it comme $\sqrt{\epsilon}$ car $\epsilon>0$, les trajectoires s'\'eloignent du foyer \`a une distance proportionnelle \`a $\sqrt{\epsilon}$ et s'enroulent autour de ce cycle limite stable. Dans ce cas la bifurcation est supercritique et donc il y a une excitation douce d'auto-oscillations. Ce résultat est justifié par le commentaire fait par V. Arnold sur le théorème de PAH dans son ouvrage [5].

Pour confirmer nos résultats, nous avons simulé l'équation de cet oscillateur en utilisant Mathématica. 


\begin{figure}[htbp]
\begin{center}
 \includegraphics[width=4cm, height=3cm]{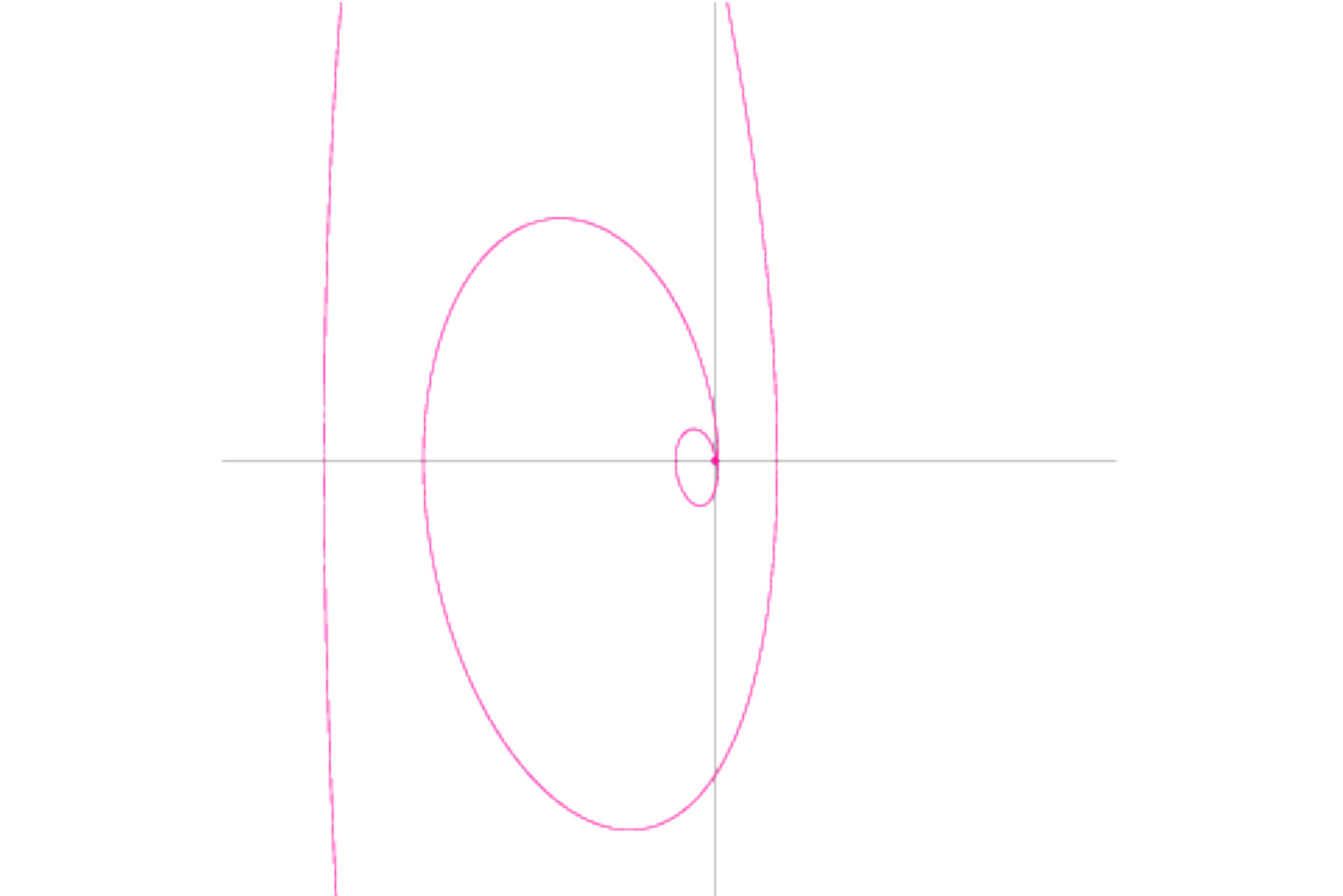}
\end{center}
\caption{ cas $\epsilon=-1.25$}
 
\end{figure}

\begin{figure}[htbp]
\begin{center}
 \includegraphics[width=6cm, height=4cm]{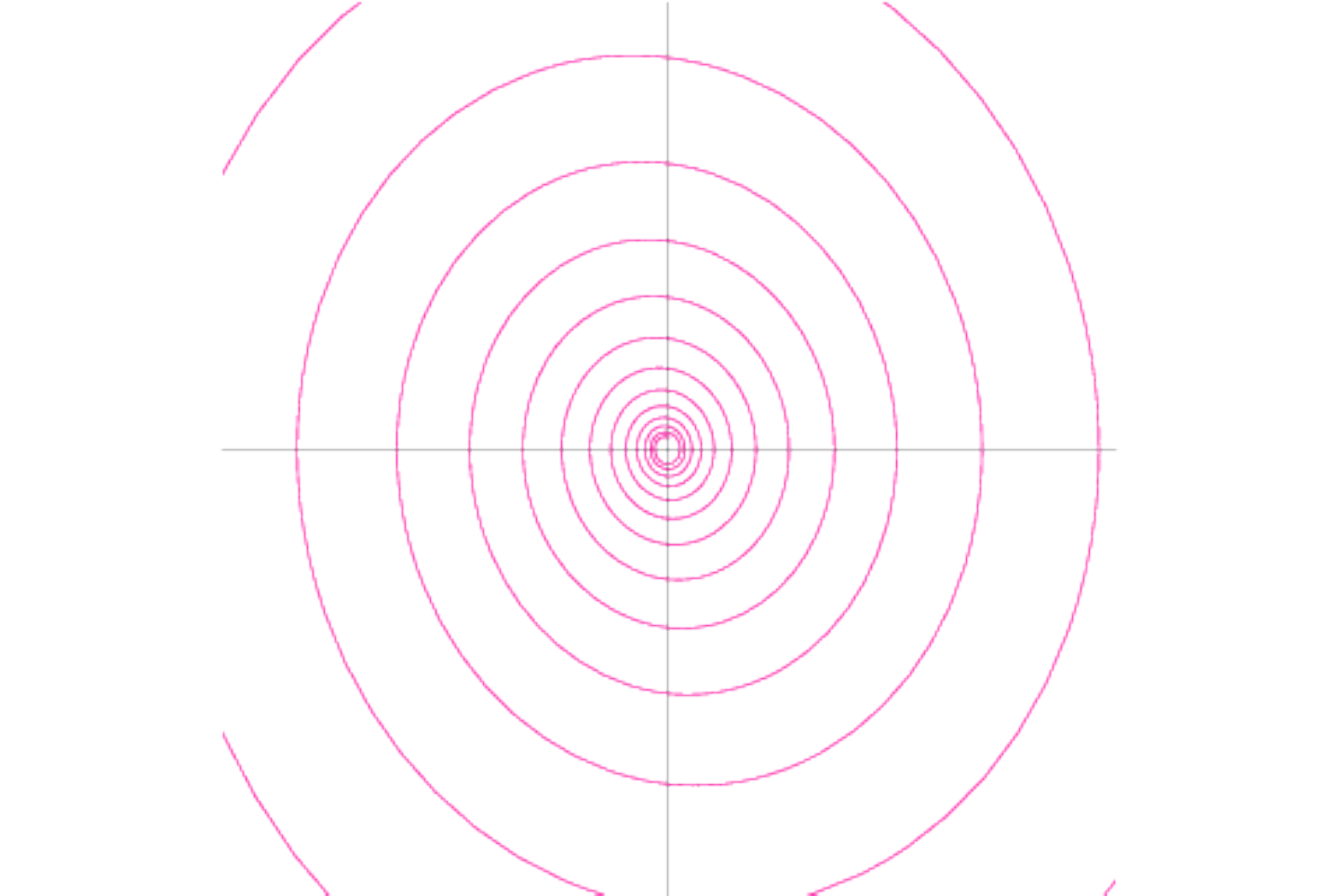}
\end{center}
\caption{cas $\epsilon=-0.25$}
 
\end{figure}

\begin{figure}[htbp]
\begin{center}
 \includegraphics[width=6cm, height=3cm]{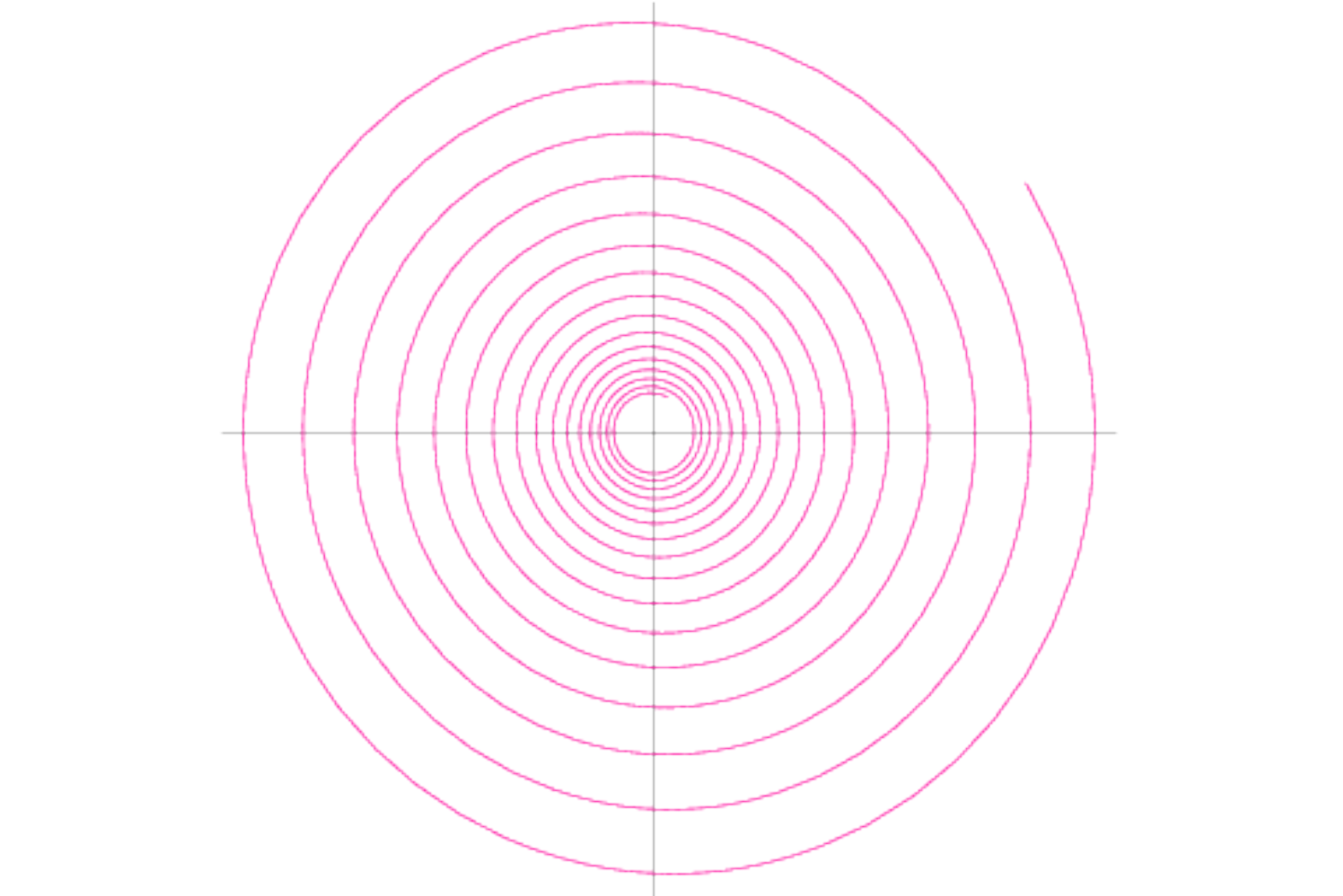}
\end{center}
\caption{ cas $\epsilon=0.0$}
 
\end{figure}

\begin{figure}[htbp]
\begin{center}
 \includegraphics[width=5cm, height=3cm]{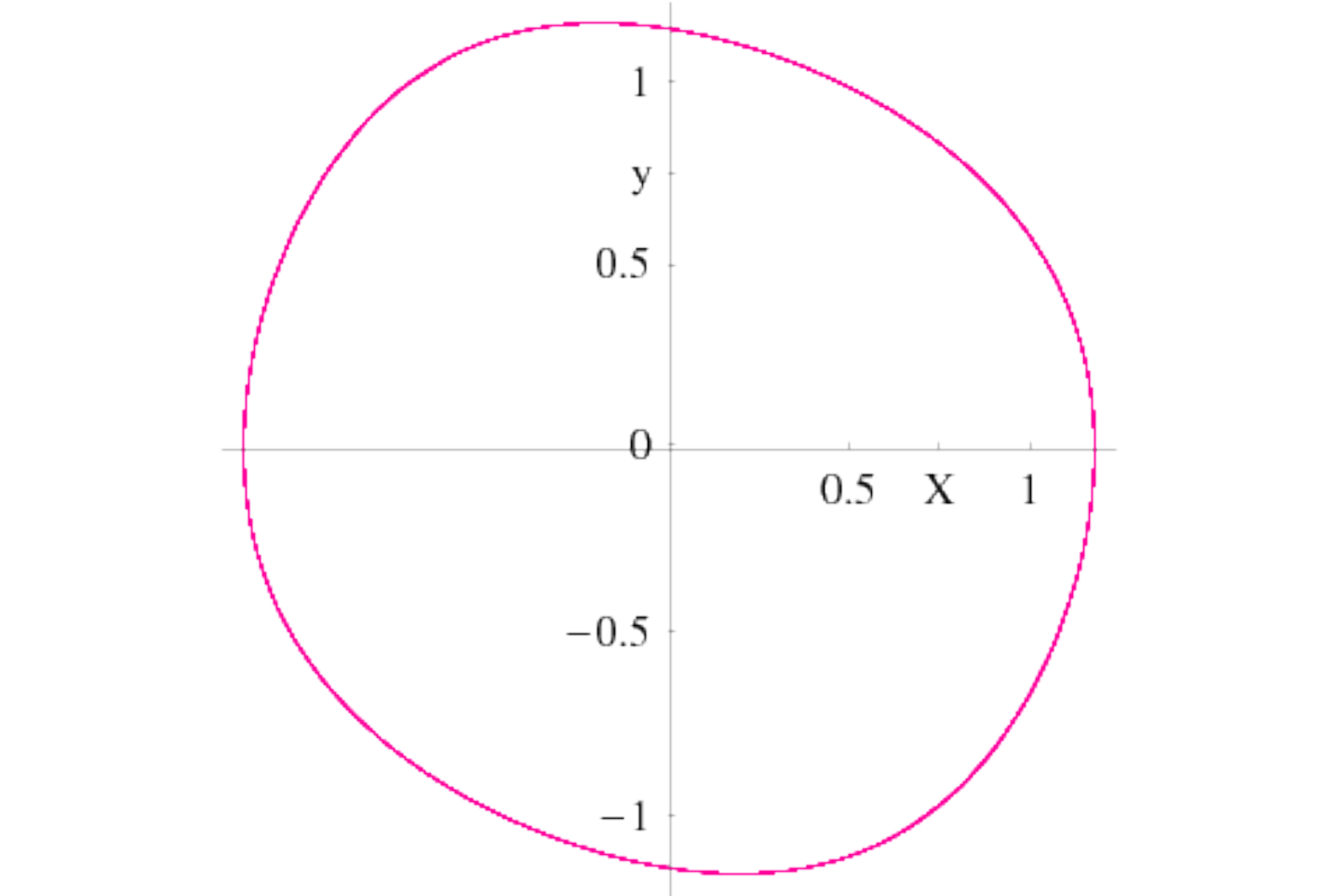}
\end{center}
\caption{ cas $\epsilon=0.5$}
 
\end{figure}

\begin{figure}[htbp]
\begin{center}
 \includegraphics[width=10cm, height=8cm]{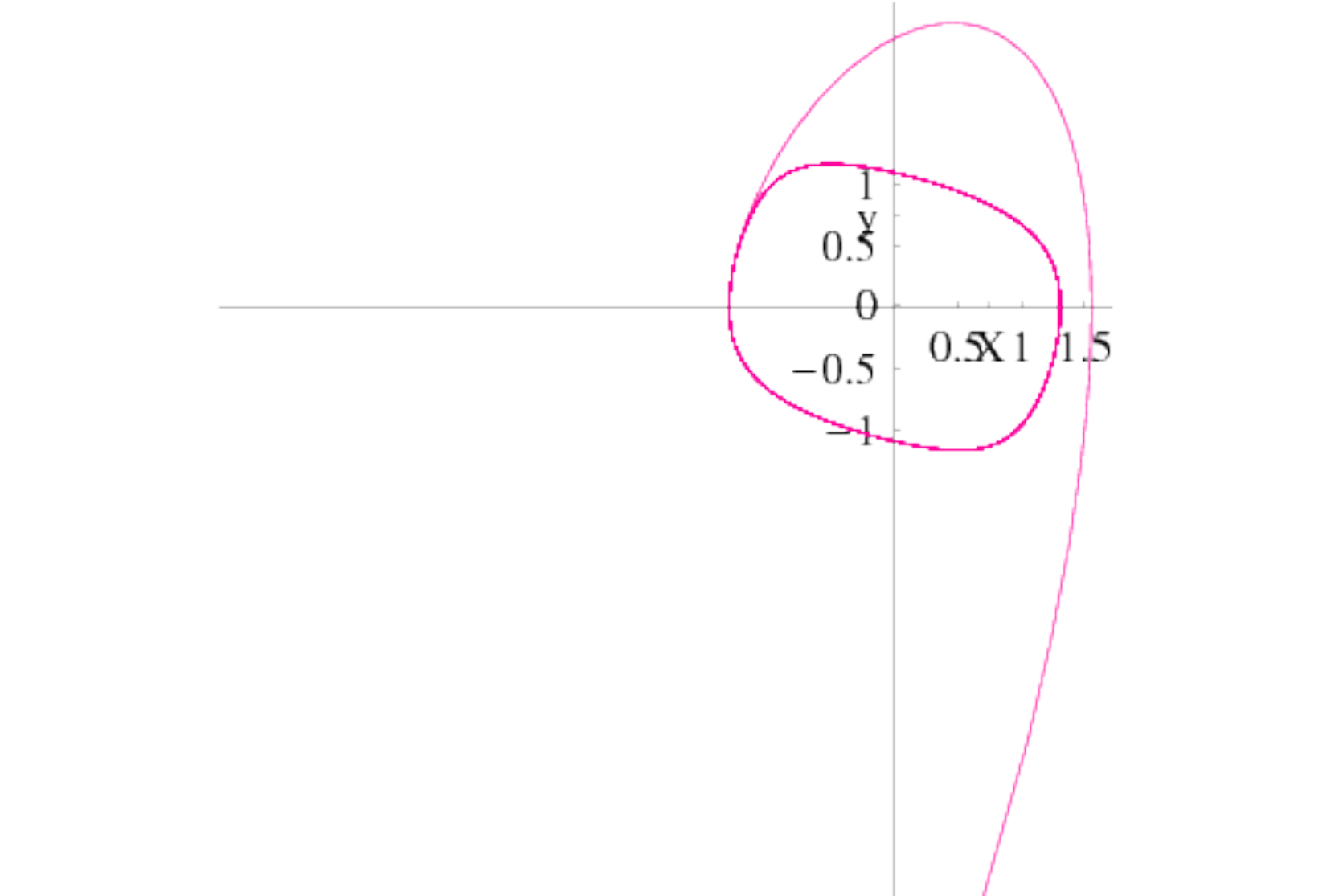}
\end{center}
\caption{ cas $0<\epsilon<2$}
 
\end{figure}

 \newpage
 
Nous remarquons que les simulations num\'eriques (figures:1,2,3,4 et 5) rendent bien compte des pr\'edictions analytiques. Nous notons aussi la pr\'esence d'un cycle limite stable et nous avons donc bien une bifurcation de Poincar\'e-Andronov-Hopf supercritique pour $\epsilon=0$. La pr\'esence de la bifurcation de PAH montre qu'un \'equilibre stable de l'oscillateur de Rayleigh se d\'estabilisera dans l'espace des param\`etres de contr\^ole en suivant le diagramme de cette bifurcation par passage d'une valeur propre \`a travers l'axe des imaginaires purs sans que le système ne présente un aspect chaotique ni catastrophique.

  \section{Conclusion}

 L'\'etude de l'oscillateur de Rayleigh nous a permis de montrer qu'il poss\`ede un cycle limite. Pour \'etudier la stabilit\'e de ce cycle limite, nous avons utilis\'e la m\'ethode de moyennage et la m\'ethode des \'echelles multiples. Il ressort de l'application de ces deux m\'ethodes que le cycle limite d'amplitude $\sqrt{\frac{4}{3}}$ de l'oscillateur de Rayleigh est stable.

Nous avons aussi, prouv\'e l'existence d'une bifurcation de Poincar\'e-Andronov-Hopf pour cet oscillateur et d\'etermin\'e analytiquement puis v\'erifi\'e par simulation num\'erique la valeur du param\`etre de contr\^ole $\epsilon$ pour laquelle cette bifurcation est obtenue. Il en r\'esulte que l'oscillateur de Rayleigh poss\`ede des propri\'et\'es qui v\'erifient les hypot\`eses de cette bifurcation. Ainsi,la présence de cette bifurcation supercritique de PAH justifie que l'oscillateur autonome de Rayleigh ne pourra pas présenter quelques soient les valeurs du paramètre de contrôle un aspect chaotique ni catastrophique.
  
\section*{Acknowlegments}
Nous remercions l'Université d'Abomey-Calavi et l'Institut de Mathématiques et de Sciences Physiques pour leurs soutients de toutes natures.

\end{document}